\documentclass[twocolumn,prl,showpacs
]{revtex4}
\usepackage{amsmath}
\usepackage{amstext}
\usepackage{latexsym}
\usepackage{psfig}
\usepackage{graphicx}
\usepackage{amsfonts}

\begin{document}
  
\title{Transfer of Nonclassical Properties from A Microscopic
Superposition to Macroscopic Thermal States in The High Temperature Limit}
  
 \author{Hyunseok Jeong} \email{jeong@physics.uq.edu.au} 
\affiliation{
Department of Physics, University of Queensland, St Lucia, Qld 4072,
   Australia }
   \author{Timothy C. Ralph}
   \affiliation{
Department of Physics, University of Queensland, St Lucia, Qld 4072,
   Australia }
  
  \date{\today}  

\begin{abstract}
We present several examples 
where prominent quantum properties 
are transferred from a microscopic superposition to thermal states
at high temperatures.
Our work is motivated by an
analogy of Schr\"odinger's
cat paradox, where the state corresponding to the virtual cat is
a mixed thermal state with a large average photon number.
Remarkably, quantum entanglement can be produced 
between thermal states with nearly
the maximum Bell-inequality violation
even when the temperatures of both modes approach infinity.
\end{abstract}


\maketitle

{\it Introduction} - 
It is a fundamental issue in quantum theory to understand 
how quantum properties such as quantum interference and
entanglement can be
transferred from microscopic objects to macroscopic classical systems.
Schr\"odinger's cat paradox is
probably the best known example of 
a counter-intuitive situation arising from the interaction
 between quantum and classical worlds \cite{Schr,LeggettReid}.
In Schr\"odinger's original paradox and its various explanations,
the initial ``cat'' put into the box is considered
to be in a pure state such as $|alive\rangle$. 
The macroscopic system,
the ``cat'', then interacts with a microscopic atomic system and
becomes entangled with it. 
According to this 
argument
a superposition 
of the macroscopic system such as
$(|alive\rangle+|dead\rangle)/\sqrt{2}$ may be
produced by measuring the microscopic part on a
superposed basis.

However, in the original gedanken experiment,
it is obvious that the macroscopic system has been 
interacting with the environment before being placed in the box. 
These interactions can cause the macroscopic system to become entangled with the 
environment before it interacts with the atomic system.
Even though the box is ideally sealed, the macroscopic system will
remain entangled with the  environment
due to its pre-interactions with the environment. 
In such a case, the macroscopic system cannot be assumed
as a pure state but it should be considered a 
mixed state \cite{Wiseman}.
Therefore, a more reasonable
assumption would be that the macroscopic system 
was initially in a mixed state before it interacted
with the microscopic superposition. 
These observations naturally lead to the
question: if the ``cat'' in Schr\"odinger's paradox was in a
mixed state, how would it be
affected by the interaction with a microscopic superposition?
Would this assumption wash out the quantum nature of the resulting state? 
This question clearly also relates more generally to the issue
of the quantum to classical interface.

In this work, we consider such an interaction between a microscopic superposition and
a significantly mixed thermal state at a high temperature.
A thermal state with a very high temperature is considered a classical state
in quantum optics. When the temperature approaches infinity, the thermal state
does not show any quantum properties.
As a comparison, coherent states  with large amplitudes are
known as the ``most classical'' pure states \cite{Schr2},
and their superposition 
is sometimes regarded as a superposition of classical states
\cite{WScat}. However, coherent states are not strictly classical
as they can be used for quantum key distribution \cite{QKD}
and display other nonclassical features
\cite{Johansen}.

Our examples in this Letter
show that prominent quantum properties can be transferred
from a microscopic superposition to significantly mixed thermal states
at high temperatures through an experimentally feasible process.
Remarkably, we find that quantum entanglement can be produced 
between thermal states with nearly
the maximum Bell-inequality violation (i.e. up to Cirel'son's bound \cite{C80})
even when the temperatures of both modes go to infinity.
To our knowledge, this interesting result has not been previously described.
For example, Ferreira {\it et al.} recently showed 
that entanglement can be generated
at any finite temperature between high Q cavity mode
field and a movable mirror thermal state \cite{FV}.
However, in their example \cite{FV}
only one of the modes is considered a large thermal state
and entanglement actually vanishes in the infinite temperature limit,
which is obviously in contrast to our result.
It is believed
that high temperatures reduce entanglement and all entanglement
vanishes if the temperature is high enough \cite{FV,Vedral}. Our result
overturns this belief and is distinguished from all the previous related works \cite{FV,Vedral,Bose,Filip}.
It also raises another interesting question on the possibility of efficient
quantum information processing with high-temperature mixed systems.

{\it Generating thermal-state superpositions} - 
Let us first consider a two-mode harmonic oscillator system.
A displaced thermal state can be defined as
$\rho^{th}(V,d)=\int d^2\alpha P^{th}(V,d)
|\alpha\rangle\langle\alpha|$
where $|\alpha\rangle$ is a coherent state of amplitude $\alpha$ and
$P^{th}(V,d)=\frac{2}{\pi(V-1)}
\exp[-\frac{2|\alpha-d|^2}{V-1}]$
with $V$ and $d$, the variance
and the displacement in the phase space, respectively.
The thermal temperature $\tau$ increases as $V$ increases
as $e^{\hbar\nu/\tau}=(V+1)/(V-1)$,
 where $\hbar$ is Planck's constant and
$\nu$ is the frequency \cite{Walls}.
Suppose that a microscopic superposition state,
$|\psi\rangle_a=(|0\rangle_a+|1\rangle_a)/2$,
where $|0\rangle$ and $|1\rangle$ are the ground and first excited states
of the harmonic oscillator,
 interacts with a thermal state $\rho^{th}_b(V,d)$ and 
the interaction Hamiltonian is
$H_\lambda= \lambda \hat a^\dagger \hat a \hat b^\dagger \hat b$
which corresponds to
the cross Kerr nonlinearity
with the nonlinear strength $\lambda$.
The resulting state is then
\begin{equation}
\begin{aligned}
\label{tmc}
&\rho^{ent}_{ab}=\frac{1}{2}
\int d^2\alpha P^{th}(V,d)
\Big\{
|0\rangle\langle 0|\otimes|\alpha\rangle\langle\alpha|+|1\rangle\langle 0| \\
&\otimes|\alpha e^{i\varphi}\rangle\langle\alpha|
+|0\rangle\langle 1|\otimes|\alpha\rangle\langle\alpha e^{i\varphi}|
+|1\rangle\langle 1|\otimes|\alpha
e^{i\varphi}\rangle\langle\alpha e^{i\varphi}|
\Big\}
\end{aligned}
\end{equation}
where $\varphi=\lambda t$ 
and $t$ is the interaction time.
\begin{figure}
\centerline{\scalebox{0.47}{\includegraphics{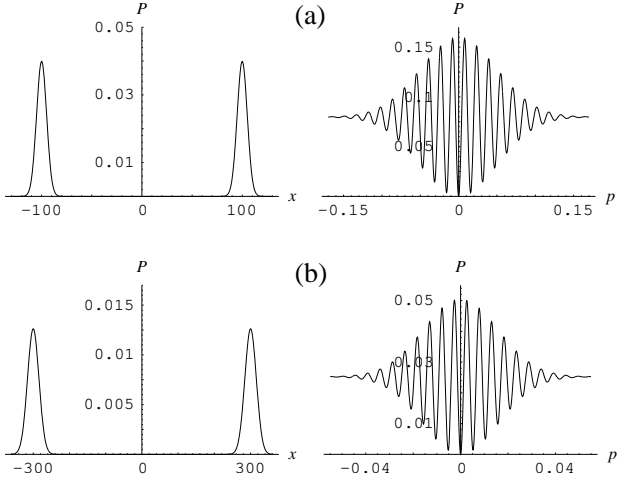}}}
\caption{The probability distributions $P$ of $x$ (left) and $p$
(right) for a ``superposition'' of 
two distant thermal states. 
A thermal state with a large mixedness 
is converted to such a ``thermal-state superposition'' 
by interacting with a microscopic superposition. 
The variance $V$ and displacement $d$
for the thermal state are chosen as (a) $V=100$ and $d=100$,
and (b) $V=1000$ and $d=300$.
The fringe visibility is 1 regardless of $V$.}
\label{fig1}
\end{figure}
The Wigner representation of $\rho^{ent}_{ab}$ is
\begin{equation}
\begin{aligned}
&W^{ent}_{ab}(\alpha,\beta)=\frac{1}{\pi}e^{-2|\alpha|^2}\Big\{
W^{th}(\beta;d)+2\alpha V^{c}(\beta;d)\\
&~~~~~+2[\alpha V^{c}(\beta;d)]^*+(4|\alpha|^2-1)W^{th}
(\beta;de^{i\varphi})\Big\}
\label{went}
\end{aligned}
\end{equation}
where $\alpha$ and $\beta$ are complex numbers parametrizing the phase spaces
of the microscopic and macroscopic systems respectively and
\begin{eqnarray}
&&W^{th}(\alpha;d)=\frac{2}{\pi V}
\exp\Big[-\frac{2|\alpha-d|^2}{V}\Big],\\
&&V^{c}(\alpha;d)=\frac{2}{\pi J K}
\exp\Big[-\frac{2}{K}(1-e^{i \varphi})d^2 \nonumber\\
&&~~~~~~~~~~~~~~~-\frac{1}{J}
(\alpha-\frac{2e^{i\varphi}d}{K})(\alpha^*-\frac{2d}{K})\Big],
\end{eqnarray}
$K=2+(V-1)(1-e^{i\varphi})$,
$J=(\sin\varphi/2+iV\cos\varphi/2)/(2V\sin\varphi/2+2i\cos\varphi/2)$,
and $d$ has been assumed real without loss of generality.
If one traces $\rho_{ab}^{ent}$ over mode $a$, the remaining state
will be simply in a classical mixture of two thermal states
and its Wigner function will be non-negative everywhere.
However, if one measures out the ``microscopic part'' 
on the superposed basis, i.e.,
$(|0\rangle_a\pm|1\rangle_a)/\sqrt{2}$,
the ``macroscopic part'' for mode $b$ 
may not lose its nonclassical characteristics.
Such a measurement on the the superposed basis 
will reduce the remaining state to
\begin{equation}
\begin{aligned}
\label{smc0}
&\rho^{sup(\pm)}=
{\cal N}_s^\pm
\int d^2\alpha P^{th}(V,d)
\Big\{
|\alpha\rangle\langle\alpha|
\pm|\alpha e^{i\varphi}\rangle\langle\alpha|\\
&~~~~~~~~~~~~~~~~~~~~~~~~~~~~
\pm|\alpha\rangle\langle\alpha e^{i\varphi}|
+|\alpha e^{i\varphi}\rangle\langle\alpha e^{i\varphi}|
\Big\},
\end{aligned}
\end{equation}
where ${\cal N}_s^\pm$ are the normalization factors,
 and its Wigner function is
$W^{sup(\pm)}(\alpha)
={\cal N}_s^\pm\{W^{th}(\alpha;d)\pm V^{c}(\alpha;d)
\pm\{V^{c}(\alpha;d)\}^*+W^{th}(\alpha;de^{i\varphi})\}$.
The $\pm$ signs correspond to the two possible results from
the measurement of the microscopic system.
The states $\rho^{sup(\pm)}$
can be understood as a generalization
of the pure superpositions of coherent states
\cite{WScat,Tu} to high-temperature thermal mixtures.

Let us first suppose that $\varphi=\pi$.
The state (\ref{smc0}) then becomes
$\rho^\pm\propto\rho^{th}(V,d)\pm\sigma(V,d)\pm\sigma(V,-d)+\rho^{th}(V,-d)$,
where 
$\sigma(V,d)=\int d^2\alpha P^{th}(V,d)|-\alpha\rangle\langle\alpha|$.
If the initial state for mode $b$ is a pure coherent state,
i.e., $V=1$, the measurement on the superposed basis 
for mode $a$ will produce 
a superposition of two pure coherent states as
$|\psi_\pm\rangle
\propto|\alpha\rangle\pm|-\alpha\rangle$.
The probability ${\cal P}_\pm$ to obtain the state $\rho^\pm$ is
${\cal P}_\pm=(1\pm\exp[-2 d^2/V])/2$.
Suppose that both $V$ and $d$ are very large for the initial thermal state.
Note that two thermal states $\rho^{th}(V,\pm d)$ become macroscopically
distinguishable when $d\gg\sqrt{V}$.
We have found that both the states $\rho^\pm$ in this case show
probability distributions with
two Gaussian peaks and interference fringes.  
Fig.~\ref{fig1}
presents the probability distributions
of $x~(\equiv Re[\alpha])$ and $p~(\equiv Im[\alpha])$
for $\rho^-$  (a) when $V=100$ and $d=100$
and  (b) when $V=1000$ and $d=300$.
The probability distribution of $x$ ($p$) 
for $\rho^\pm$ can be obtained by
integrating the Wigner function of $\rho^\pm$ over $p$ ($x$).    
The two Gaussian peaks along the $x$ axis and
interference fringes along the $p$ axis shown in Fig.~\ref{fig1}
are a typical signature of a quantum superposition between 
macroscopically distinguishable states.
The visibility $v$ of the interference fringes is defined as
\cite{Walls}
$v=(I_{\max}-I_{\min})/(I_{\max}+I_{\min})$,
where $I=\int dx W^{sup(-)}(\alpha)$ and the maximum and minimum
should be taken over $p$.
It can be simply shown that the visibility $v$ can be always 1 
regardless of the value of $V$. 
Note that $d$ should increase proportionally to $\sqrt{V}$ 
to maintain the condition of classical distinguishability between
the two component thermal states $\rho^{th}(V,\pm d)$.
\begin{figure}
\centerline{\scalebox{0.28}{\includegraphics{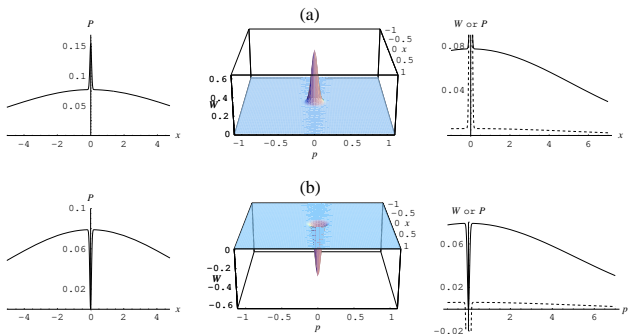}}}
\caption{The probability distributions $P$ (left) and the Wigner functions
$W$ (middle) of the thermal state of $V=100$
around the origin ($d=1$) after an interaction with
a microscopic superposition and a conditional measurement.
The right-hand-side figures show
that the background of the probability distributions (solid line) 
relates to the nonzero background of the Wigner functions (dashed line)
around the sharp pole and the hole.
}
\label{fig2}
\end{figure}
The interference fringes with high visibility
are incompatible with classical physics
and evidence of quantum coherence.
The fringe spacing (the distance between the fringes)
does not depend on $V$ but only on $d$.
A pure superposition of coherent states shows
the same fringe spacing as the mixed ones
in Fig.~\ref{fig1} for a given $d$. 
We emphasize that the states shown in Fig.~\ref{fig1} are
``superpositions'' of severely mixed thermal states.

When $d$ is small while $V$ is still very large, i.e.,
a classical thermal state 
around the origin is assumed as the initial state,
the resulting thermal-superposition states
also show peculiar nonclassical behaviors.
Fig.~\ref{fig2} shows the probability distribution and
the Wigner functions of the
states (a) $\rho^+$ and (b) $\rho^-$
where $V=100$ and $d=1$.
As a result of the interaction with the microscopic
superposition, a deep hole 
has been formed around the origin for $\rho^-$ 
(or a sharp pole toward the top for the case of $\rho^+$).
Interestingly, the Wigner function of $\rho^-$ in Fig.~\ref{fig2}(b) 
has a deep hole to the negative direction below zero. 
In this case, however, the probability ${\cal P}_-$ 
is only 1\% while ${\cal P}_+$ is 99\%.

An experimental realization of a nonlinear effect
corresponding to $\varphi=\pi$ is very demanding
particularly in the presence of decoherence.
Here we point out that 
the method using a weak nonlinear effect
($\varphi\ll \pi$)
combined with a strong field 
($d\gg 1$) \cite{Jeong05} can
be useful to generate a thermal-state superposition
with prominent interference patterns.
In Fig.~3, we have used experimentally accessible values, $V=5$, $d=2000$ and 
$\varphi=\pi/1000$, but the fringe visibility is still 1.
In this case, {\it decoherence during the nonlinear interaction
would be significantly reduced} because of 
the decrease of the interaction time
\cite{Jeong05}.
Note also that, if required, the state in Fig.~3 
can be moved to the center of the phase space,
for example, using a biased  beam splitter (BS)
and a strong coherent field
\cite{Jeong05}.

\begin{figure}
\centerline{\scalebox{0.45}{\includegraphics{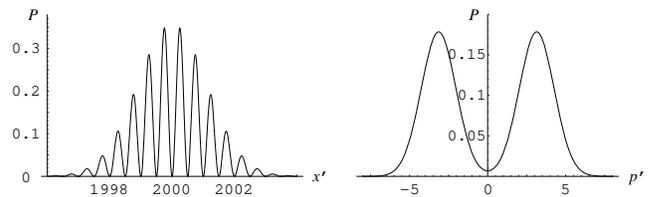}}}
\caption{The probability distributions $P$ 
for a ``superposition'' of thermal states where  $V=5$, $d=2000$,
$\varphi=\pi/1000$. 
The $x^\prime$ ($p^\prime$) axis 
in this figure has been rotated  by $\pi/2000$ from
the $x$ ($p$) axis for clarity.}
\label{fig4}
\end{figure}
 
\begin{figure}
\centerline{\scalebox{0.45}{\includegraphics{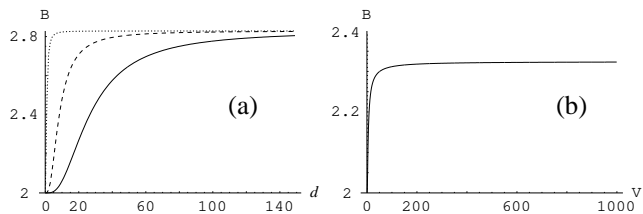}}}
\caption{(a) The optimized violation, ${\rm B}\equiv|B^+|_{max}$,
of the Bell-CHSH inequality for the
``thermal-state entanglement'', $\rho^{tm(+)}$, of $V=1000$ (solid line)
and $V=100$ (dashed line). The Bell-violation of a pure entangled
coherent state, i.e., $V=1$, has been plotted for comparison (dotted line).
(b) The optimized Bell-violation B against $V$ for the slightly different
type of thermal-state entanglement
generated using  a 50:50 BS when $d=0$. See text for details.}
\label{bell}
\end{figure}

{\it Entanglement between thermal states} -
There remains an important question concerning the possibility of generating 
entanglement between macroscopic objects and 
its Bell-type inequality tests \cite{Bell}.
We shall show that entanglement can be generated
between high-temperature thermal states
even when the temperature of each mode goes to infinity.
If the microscopic superposition interacts with two thermal states,
$\rho_b^{th}(V,d)$ and $\rho_c^{th}(V,d)$,
and the microscopic particle is measured out on the superposed basis
\cite{mskim}, the resulting state will be
\begin{eqnarray}
\begin{aligned}
&&\rho^{tm(\pm)}
\propto\rho^{th}(V,d)\otimes\rho^{th}(V,d)
\pm\sigma(V,d)\otimes\sigma(V,d)~~~\\
&&\pm\sigma(V,-d)\otimes
\sigma(V,-d)+\rho^{th}(V,-d)\otimes\rho^{th}(V,-d).
\end{aligned}
\label{e-tm}
\end{eqnarray}
Thermal fields in two cavities and a two-level atom  
can be used to generate such a state.
Here we shall consider the Bell-CHSH inequality \cite{Bell,CHSH} 
with photon number parity measurements \cite{mskim,BW}. 
Such parity measurements can be made in
a high-Q cavity using a far-off-resonant
interaction between a two-level atom and the field \cite{eh}.
The Bell-CHSH inequality can be
represented in terms of the Wigner function as
\cite{BW}
\begin{equation}
\begin{aligned}
&|B^{(\pm)}|=\frac{\pi^2}{4}\big|W^{tm(\pm)}(\alpha,\beta)
+W^{tm(\pm)}(\alpha,\beta^\prime)\\
&~~~~~~~~~~~+W^{tm(\pm)}(\alpha^\prime,\beta)
-W^{tm(\pm)}(\alpha^\prime,\beta^\prime)\big|\leq2,
\end{aligned}
\end{equation}
where $W^{tm(\pm)}(\alpha,\beta)$ is the Wigner function of
$\rho^{tm(\pm)}$ in Eq.~(\ref{e-tm}).
As shown in Fig.~\ref{bell}, the Bell-violation
approaches the maximum bound for a bipartite measurement,
$2\sqrt{2}$ \cite{C80}, when $d\gg\sqrt{V}$ 
regardless of the level of the mixedness $V$,
i.e., the temperatures of the thermal states.
Note that it is true for both of $\rho^{tm(+)}$ and $\rho^{tm(-)}$
even though only the case
of $\rho^{tm(+)}$ has been plotted in Fig.~\ref{bell}(a). 
This implies that entanglement of nearly 1 ebit has been produced 
between the two significantly mixed thermal states for
$d\gg\sqrt{V}$, and such ``thermal-state entanglement''
cannot be described by a local theory.

Slightly different types of macroscopic 
entanglement can be generated
by applying the BS operation on 
the ``thermal-state superpositions'' in Eq.~(\ref{smc0}).
When $d$ is large, these states, which are generated by a 50:50 BS, 
violate the Bell-CHSH inequality to the maximum bound $2\sqrt{2}$
regardless of the level of mixedness $V$ in exactly the same way described above.
Furthermore, these states severely violate Bell's inequality even when $d=0$ as
$V$ increases as shown in Fig.~\ref{bell}(b).
We have found that the optimized Bell violation of these states
approaches $2.32449$ for $V\rightarrow\infty$.
Interestingly, this value is exactly the same as the optimized  Bell-CHSH violation
for a pure two-mode squeezed state in the infinite squeezing limit \cite{jeongsonkim}.
This is a remarkable distinguishing feature of
the generation process of the thermal-state superpositions from
that of the pure coherent-state superpositions
using an initial coherent state \cite{Tu}.
If a coherent state with $d=0$ (i.e. the vacuum) was
used, the initial vacuum state
would remain the same and no quantum effects described above can manifest.

{\it Experimental feasibility} -
A feasible experimental setup
to test our examples is atom-field interactions in cavities
where a $\pi/2$ pulse can be used to prepare the atom in a superposed state
\cite{Tu}.
The two-mode thermal-state entanglement can be generated using
two cavities and an atomic state detector \cite{mskim}.
Extending the two cavities to $N$, 
 entanglement of
$N$-mode thermal states can also be generated.
Another possible setup is
an all-optical scheme with free-traveling fields and a cross-Kerr medium,
where a standard single-photon qubit
could be used as the microscopic superposition \cite{Jeong05}.
Entanglement in a traveling field configuration can be
produced at a BS.
The observation of interference fringes can be performed
using homodyne detection, which is generally very efficient,
in a very short time. 

Finally, we assess decoherence effects
on quantum nonlocality (i.e., Bell-inequality violations)
in a feasible range for 
photon number parity measurements.
The decoherence analysis can be performed 
 using the known solution of the master equation
 for a coherent-state dyadic:
$|\alpha\rangle\langle\beta|\rightarrow
e^
{-(1-e^{-\gamma t})\{(|\alpha|^2+|\beta|^2)/2-\alpha\beta^*\}}
|e^{-\gamma t/2} \alpha \rangle\langle e^{-\gamma t/2}\beta |$,
where $\gamma$ is the energy decay rate
and $t$ is time 
\cite{Jeong05}.
As can be expected, the increase of either $V$ or $d$
causes a more rapid destruction of nonlocality. However,
when $V$ takes moderate values, the condition for nonlocality
($|B^{\pm}|_{max}>2$)
can persist
fairly long compared to typically discussed pure cases.
We suggest an experimentally feasible case, i.e.,
$V=3$ and $d=1$. In this case, the degree of mixedness 
of the initial thermal state in terms of
the linear entropy ($M=1-{\rm Tr}[\rho^2]$) is $\approx 0.67$, i.e.,
 a very mixed state. However, nonlocality
of a generated state, $\rho^-$,
divided at a 50:50 BS persists slightly
longer (until $\gamma t\approx0.13$) than a pure superposition state, $|\alpha\rangle+|-\alpha\rangle$,
with $\alpha=2.2$ divided at a 50:50 BS (until $\gamma t\approx0.12$).
Here, the total average photon number for $\rho^-$
(at $t=0$) is $\approx2.5$,
i.e., the average photon number detected at each detector is only $\approx1.3$.
Then, there is a good possibility of performing
parity measurements using current technology.
As another case, if $V=10$ (then $M=0.9$)
and $d=0$, the survival time of nonlocality
for the generated state $\rho^+$
divided at a 50:50 BS is
approximately the same as that of 
a pure coherent-state superposition with $\alpha=3.55$ (until $\gamma t\approx0.05$).
The average photon number for {\it each} detector in this case 
is $\approx2.0$. 
In summary, we have shown that superpositions of ``classical'' thermal states
still show strong quantum effects, even when temperatures approach infinity, 
and that their small scale demonstrations at fairly low temperatures
appear experimentally feasible.

This work was supported by the Australian Research Council 
and the University of Queensland.
H.J. thanks M.S. Kim, M. Paternostro, P.L. Knight, J. Lee and
R. Filip for discussions and feedback.


\begin{references} 

\bibitem{Schr} 
E. Schr$\rm\ddot{o}$dinger, {\it Naturwissenschaften.} {\bf 23},
 pp. 807-812; 823-828; 844-849 
(1935).

\bibitem{LeggettReid} A.J. Leggett and A. Garg,  Phys.Rev.Lett.
{\bf 54}, 857 (1985);
M.D. Reid, quant-ph/0101052 and references therein.

\bibitem{Wiseman} H.M. Wiseman and J.A. Vaccaro, Phys.Rev.Lett. {\bf 87},
240402, (2001); See discussions in the introduction and references therein.

\bibitem{Schr2} 
E. Schr$\rm\ddot{o}$dinger, {\it Naturwissenschaften.} {\bf 14}, 664
(1926).

\bibitem{WScat} W. Schleich, M. Pernigo, and F.L. Kien,
Phys. Rev. A {\bf 44}, 2172 (1991).

\bibitem{QKD} F. Grosshans {\it et al.}, Nature (London) 421, 238 (2003).

\bibitem{Johansen} L.M. Johansen, Phys. Lett. A {\bf 329}, 184 (2004).

\bibitem{C80} B.S. Cirel'son, Lett. Math. Phys. {\bf 4}, 93 (1980).

\bibitem{FV}  A. Ferreira, A. Guerreiro, and V. Vedral, 
Phys. Rev.  Lett. {\bf 96}, 060407 (2006).

\bibitem{Vedral} V. Vedral, New J. Phys. {\bf 6} 102 (2004).

\bibitem{Bose} S. Bose, 
 I. Fuentes-Guridi, P.L. Knight, and
V. Vedral,
 Phys.Rev.Lett. 87, 050401 (2001).

\bibitem{Filip} R. Filip, 
 M. Du\v sek, J. Fiur\' a\v sek, and L. Mi\v sta,
Phys.Rev.A {\bf 65}, 043802 (2002).

\bibitem{Tu} M. Brune {\it et al.}, Phys.Rev.Lett. {\bf 77},
  4887 (1996); A. Auffeves {\it et al.},
 Phys.Rev.Lett. {\bf 91} 230405 (2003).

\bibitem{Walls} D.F. Walls and G.J. Milburn, {\it Quantum Optics}, Springer-Verlag (1994).

\bibitem{Jeong05} H. Jeong, Phys.Rev.A {\bf 72}, 034305 (2005) and references therein.

\bibitem{Bell}  J.S. Bell, Physics  (Long Island City, N.Y.)  {\bf 1}, 195 (1964).

\bibitem{mskim} M.S. Kim and J. Lee, Phys.Rev.A {\bf 61} 042102 (2000).

\bibitem{CHSH}  J.F. Clauser {\it et al.},
  Phys.Rev.Lett. {\bf 23}, 880 (1969).
 
\bibitem{BW} K. Banaszek and K. W{\' o}dkiewicz, Phys.Rev.A {\bf 58},
4345 (1998).

\bibitem{eh} B.-G. Englert, N. Sterpi, and H. Walther, Opt. Commun.
{\bf 100} 526 (1993).
 
\bibitem{jeongsonkim} H. Jeong {\it et al.},
Phys.Rev.A {\bf 67}, 012106 (2003).




\end{references}
\end{document}